\documentclass[multphys,vecphys]{svmult}

\usepackage{makeidx}     % allows index generation
\usepackage{graphicx}    % standard LaTeX graphics tool
\usepackage{multicol}    % used for the two-column index
\makeindex             % used for the subject index
\def\arcsec{\hbox{$^{\prime\prime}$}}
\begin{document}

\title*{High-angular resolution HC$_3$N and CS observations of the dense core
in the cirrus cloud MCLD123.5+24.9}
\titlerunning{A dense core in the cirrus cloud MCLD123.5+24.9}
%for an abbreviated version of
% your contribution title if the original one is too long
\author{Christoph B\"ottner \inst{1}\and
Andreas Heithausen \inst{1} \and Fabian Walter \inst{2} }
\authorrunning{C. B\"ottner, A. Heithausen, \& F. Walter}
%for an abbreviated version of
% your contribution title if the original one is too long
\institute{
%Institute for Radio Astronomy, University of Bonn, Auf dem H\"ugel 71, 53121 Bonn
Radioastronomisches Institut der Universit\"at Bonn,\\ Auf dem H\"ugel 71, 53121 Bonn\\
\texttt{cboettne@astro.uni-bonn.de},
%\and  Radioastronomisches Institut der Universitaet Bonn, Auf dem H\"ugel 71, 53121 Bonn
\texttt{heith@astro.uni-bonn.de}
\and  National Radio Astronomy Observatory, P.O. Box O, Socorro, NM, 87801, USA \texttt{fwalter@nrao.edu}
}
%
% Use the package "url.sty" to avoid
% problems with special characters
% used in your e-mail or web address
%
\maketitle
%Your text goes here. Separate text sections with the standard \LaTeX\
%sectioning commands.

{\bf Abstract:} {\small We report on high-angular resolution observations of a dense core in the cirrus cloud MCLD123.5+24.9 obtained with the PdB and the OVRO interferometers. Our maps show substructures down to the scale of the beam ($\sim 1000$\,AU). The chosen molecules CS\,(2-1) and HC$_3$N\,(10-9) trace different regions in the core. This can be explained by time-dependent chemical evolution of the cloud and therefore provides constraints on the timescales of the fragmentation and core formation processes. Our data demonstrate that the chemical evolution plays the decisive role in our attempt to interpret observational data of different gas density tracers.}

\section{Introduction}

Galactic cirrus clouds are typically complexes of diffuse gas and dust most easily seen at high galactic latitudes. On large scales they are dominated by turbulent motions, however, on small scales dense cores have been found and there is evidence that these cores are gravitationally bound (Heithausen et al. 2002).
%\cite{heith02}
In one core, even the spectral signature of infall motion was discovered (Heithausen 1999).\\
%\cite{heith99}
Core formation constitutes the very first step which leads to star-formation and is still rather poorly understood. The knowledge of how cores form and how they evolve is important for understanding the minimal physical conditions needed for stars to form. Studying the evolution of dense cores in cirrus clouds, which live in a rather quiescent environment and are mainly heated by the interstellar radiation field, may prove to be crucial to interpret observations of star-forming regions correctly.\\
One of the best studied cirrus cloud cores so far is the one in MCLD123.5+24.9, located in the Polaris Flare, a huge cirrus cloud complex in the direction of the north celestial pole. Its distance is between 130 and 240\,pc (cf. Heithausen et al. 1993),
%\cite{heith93}
and here we adopt a value of 150\,pc. The temperature of the dust was determined to 13\,K (Bernard et al. 1999)
%\cite{bern99}
and the kinetic temperatures of the gas between 6 and 15\,K  (e.g. Gro\ss mann \& Heithausen 1992).\\
%\cite{gro92}
Here we present new high-angular resolution HC$_3$N and CS observations of this core
%in MCLD123.5+24.9
which allow a much closer insight into the fragmentation and core formation processes.

\section{Observations}

The observations of the HC$_3$N\,(10-9) transition at 90.98 GHz were conducted between May and November 2002 with the IRAM Plateau de Bure (PdB) interferometer. %During four runs five antennas in the configuration 5D were used, one run was carried out with 6 antennas in configuration 6Dp. %Because of poor weather conditions only the 3mm data were usable and
The standard CLIC reduction procedure was applied. Phase stability was checked with frequent observations of the quasars 1928+738 and 0716+714 and the amplitude scale was derived from measurements of MWC349 and CRL618.\\
The CS\,(2-1) line observations were carried out using the Caltech OVRO millimeter interferometer from January to April 2001. In total, 5 tracks were spent on source in the 'C' and 'L' configurations. The nearby source J1803+784 was used for phase calibration. The data for each array were edited and calibrated separately with the MMA and MIRIAD packages. The uv-data were inspected and bad data points due to either interference or shadowing between telescopes were removed, after which the data were calibrated.\\
We then corrected the data for zero spacing with our dataset obtained with the IRAM 30m telescope. The effective synthesized beam FWHM of $\approx 6\arcsec \times 8\arcsec$ is similar in both observations (but with different position angles of $\rm{PA}=-72^{\circ}$ and $\rm{PA}=-13.5^{\circ}$, respectively) equivalent to 900\,AU $\times$ 1200\,AU at the distance of 150\,pc.
%Based on observations carried out with the IRAM Plateau de Bure Interferometer. IRAM is supported by INSU/CNRS (France), MPG (Germany) and IGN (Spain).

\section{Results}

Our maps (see Fig. 1) show substructures down to the scale of the beam and the cores are clearly elongated. Most interesting is the complete misalignment of the HC$_3$N and the CS clumps, which was already suggested by the single dish data. However, the HC$_3$N clump was not resolved at all by the single dish observations and also the CS structures were barely resolved.
The HC$_3$N clump resides at the southern edge of the core, is elongated and divides in substructure, whereas the CS emission is much more extended and the maximum, where the spectral signature of infall motion can be seen (cf. Heithausen 1999),
%\cite{heith99}
lies more towards the north of the core. Nevertheless, there is still CS emission seen along with the HC$_3$N, even a small local maximum at the position of the HC$_3$N peak.\\
Heithausen et al. (2002)
%(\cite{heith02})
presented a dust continuum emission map of the core obtained with the MAMBO array at the IRAM 30m telescope with a resolution of 11\arcsec, hence roughly comparable to our resolution. The dust shows again a somewhat different behavior than the gas components. The filamentary structure of the whole core can be seen much better and only one elongated dense core, nevertheless with a lot of substructure down to the scale of the beam, is visible. This dust core corresponds well with the CS emission but is even more extended and clearly includes also the HC$_3$N clump. Therefore we can conclude that we are looking at one cirrus cloud core with a lot of substructure due to fragmentation and strong small scale variations in the molecular abundances.

\begin{figure}
\centering
\rotatebox{-90}{\includegraphics[height=11cm]{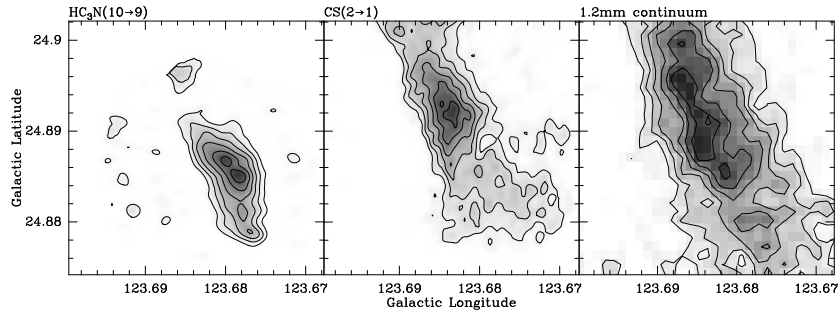}}
% If not, use
%\picplace{5cm}{2cm} % Give the correct figure height and width in cm
\caption{Maps of the core in MCLD123.5+24.9 in HC$_3$N\,(10-9), CS\,(2-1) and the dust continuum at 1.2mm.
%(from Heithausen et al. 2002) \cite{heith02}
The most striking feature is the displacement of the two molecular clumps, which are both embedded in the dust core.
%The spectral line maps are our new interferometric observations including zero spacing corrections.
}
\label{fig:1}       % Give a unique label
\end{figure}

\section{Discussion}

The most likely explanation for this appearance is that it is caused by strong abundance variations of the observed molecules due to time dependent chemical evolution. There are several chemical models in the literature which can account for the observed behavior.
One that matches our derived properties of the core quite good is the one of Markwick et al. (2000),
%(\cite{mark00})
constructed for TMC-1, a low-mass star forming region.
The initial conditions in this model are a temperature of 10\,K, a hydrogen number density $\rm{n}(\rm{H}_2)=2\times 10^{4}\rm{cm}^{-3}$, a cosmic ray ionization rate of $1 \times 10^{-17} s^{-1}$, and the UMIST Database for Astrochemistry 1995 reaction set (Millar et al. 1995).
%\cite{mill95}
This model predicts a time delay of HC$_3$N compared to CS in reaching a measurable high abundance of $(3-10) \times 10^5$\,years (see Fig. 2). By this time the CS, which attained its highest abundance very fast after only $10^3$\,years, may already start to deplete due to freezing out on dust grains.

\begin{figure}
\centering
%\rotatebox{-90}
{\includegraphics[height=4cm]{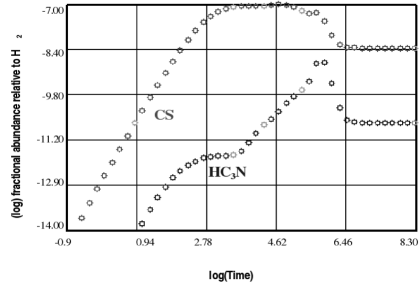}}
% If not, use
%\picplace{5cm}{2cm} % Give the correct figure height and width in cm
\caption{Abundances of CS and HC$_3$N relative to H$_2$ with time in the chemical model by Markwick et al. (2000). HC$_3$N needs about $3\times 10^5$\,years more than CS to reach a measurable high abundance. (from www.astrochemistry.net)}
\label{fig:2}       % Give a unique label
\end{figure}

This result is further supported by a time-dependent chemical model of core formation by Taylor et al. (1998),
%(\cite{tayl98})
which basically comes to the same results with somewhat different initial conditions. Their main conclusion is, that CS can be seen in more or less the whole core, whereas the HC$_3$N on the other hand can only be observed in longer-lasting clumps. This behavior is essentially a result of the cold ($\sim 10$\,K) gas chemistry where also neutral-neutral reactions play a surprisingly important role.

\section{Conclusions}

We observed a core in the cirrus cloud MCLD123.5+24.9 with high-angular resolution in the molecules HC$_3$N and CS. Our data show a spatial offset of the maxima of both molecules and substructures down to the scale of 1000\,AU. The misalignment can be explained by the interplay of time-dependent chemical evolution and fragmentation. The HC$_3$N traces only the older part of the core, whereas the CS is much closer tracing the density structure of the whole core as seen also in the dust continuum. This provides us with constraints on the timescales of the core formation process of a few $10^5$ years, the time HC$_3$N needs to reach a measurable high abundance.

%%%%%%%%%%%%%%%%%%%%%%%% referenc.tex %%%%%%%%%%%%%%%%%%%%%%%%%%%%%%
% sample references
% "physics"
%
% Use this file as a template for your own input.
%
%%%%%%%%%%%%%%%%%%%%%%%% Springer-Verlag %%%%%%%%%%%%%%%%%%%%%%%%%%

%
% BibTeX users please use
% \bibliographystyle{}
% \bibliography{}

\begin{thebibliography}{99.}
% and use \bibitem to create references.
% Use the following syntax and markup for your references
% Monographs
%\bibitem{monograph} H. Ibach, H. L\"uth: \textit{Solid-State
%Physics}, 2nd edn (Springer, Berlin Heidelberg New York 1996) pp 45--56
% Contributed Works
%\bibitem{contribution} D.M. MacKay: Visual stability and voluntary eye
%movements. In: \textit{Handbook of Sensory Physiology}, vol 3, ed by R.
%Jung, D.M. MacKay (Springer, Berlin Heidelberg New York 1973) pp
%307--331
% Journal
%\bibitem{journal} S. Preuss, A. Demchuk Jr, M. Stuke et al: Appl. Phys.
%A \textbf{61}, 33 (1995)

\bibitem{bern99} J.P. Bernard, A. Abergel, I. Ristorcelli, et al.: A\&A \textbf{347}, 640 (1999)
\bibitem{gro92} V. Gro\ss mann \& A. Heithausen: A\&A \textbf{264}, 195 (1992)
\bibitem{heith93} A. Heithausen, et al: A\&A \textbf{268}, 265 (1993)
\bibitem{heith99} A. Heithausen: A\&A \textbf{349}, L53 (1999)
\bibitem{heith02} A. Heithausen, F. Bertoldi, \& F. Bensch: A\&A \textbf{383}, 591 (2002)
\bibitem{mark00} A.J. Markwick, T.J. Millar, \& S.B. Charnley: ApJ \textbf{535}, 256 (2000)
\bibitem{mill95} T.J. Millar, P.R.A. Farquhar, \& K. Willacy: A\&AS \textbf{121}, 139 (1995)
\bibitem{tayl98} S.D. Taylor, O. Morata, \& D.A. Williams: A\&A \textbf{336}, 309 (1998)

% Theses
%\bibitem{thesis} D.W.  Ross: Lysosomes and storage diseases. MA
%Thesis, Columbia University, New York (1977)
\end{thebibliography}
%
% Non-BibTeX users please use

%%%%%%%%%%%%%%%%%%%%%%%%%%%%%%%%%%%%%%%%%%%%%%%%%%%%%%%%%%%%%%%%%%%%%%  }

%%%%%%%%%%%%%%%%%%%%%%%%%%%%%%%%%%%%%%%%%%%%%%%%%%%%%%%%%%%%%%%%%%%%%%

%\printindex
\end{document}